\documentclass[useAMS,usenatbib]{mnras}
\usepackage{graphicx}
\usepackage{mathtools}
\usepackage{gensymb}
\usepackage{aas_macros}
\usepackage{amsmath}
\usepackage[flushleft]{threeparttable}

\title{SAFARI I: A SPHERE discovery of a super metal-rich M dwarf companion to the star HD 86006}
\author[B. M. Pantoja et al.]
  {B. M.~Pantoja,$^{1,2}$ %
  J. S.~Jenkins$^{1,3}$,  J. H.~Girard$^{2,6}$, A.~Vigan$^4$, G. S.~Salter$^4$ 
 \and
 and M. I. ~Jones$^{2,5}$ \\
  $^1$ Departamento de Astronom\'{i}a, Universidad de Chile, Camino el Observatorio 1515, Las Condes, Santiago, Casilla 36-D, Chile \\
  $^2$ European Southern Observatory, Alonso de Cordova 3107, Vitacura, Santiago, Chile \\
  $^3$ Centro de Astrof\'isica y Tecnolog\'ias Afines (CATA), Casilla 36-D, Santiago, Chile \\
  $^4$ Aix Marseille Univ, CNRS, LAM, Laboratoire d'Astrophysique de Marseille, Marseille, France \\
  $^5$ Departamento de Astronom\'{i}a y Astrof\'{i}sica, Pontificia Universidad Cat\'{o}lica de Chile, Vicu\~{n}a Mackenna 4860, 7820436 Macul,\\ Santiago, Chile\\
  $^6$ Space Telescope Science Institute, Baltimore, MD, USA
}
\date{Accepted 2018 June 20. Received 2018 June 19; in original form 2017 October 05
}

\pagerange{\pageref{firstpage}--\pageref{lastpage}} \pubyear{2002}

\def\LaTeX{L\kern-.36em\raise.3ex\hbox{a}\kern-.15em
    T\kern-.1667em\lower.7ex\hbox{E}\kern-.125emX}

\begin{document}

\label{firstpage}

\maketitle 

\begin{abstract}
We report the direct detection of a fully convective, early-to-mid M-dwarf companion orbiting the star HD 86006, using ESO-SPHERE during Science Verification as part of the SAFARI program. HARPS+CORALIE radial velocity measurements first indicated a possible companion. Such work highlights the synergies that are now possible between these two observing methods. We studied the companion by comparing our observed spectra with BT-Settl models and template spectra, measuring spectral indices to obtain a spectral type, and used a joint radial velocity and astrometric fit to simulate the companion's orbit. The companion was found to be 4.14 mag fainter than the primary in the H2 band, residing at a physical separation of $\sim$ 25 AU, 
with a $T_\mathrm{eff}$ and {spectral type} of 3321 $\pm$ 111 K and {M 4.1 $\pm$ 1.1,}
respectively.  We note that the age derived from BT-Settl models for such a star is too low by over two orders of magnitude, similar to other known field mid-M stars.  We searched for the radial velocity companion to HD 90520 without any clear detection, however we reached a low contrast level of $\Delta$H2 = 10.3 mag (or $1.3 * 10^{-4}$) at 0.2$''$ and 12.6 mag (or $10^{-5}$) at 0.5$''$, allowing us to rule out any low-mass companions with masses of 0.07 and 0.05 M$_{\odot}$ at these separations. This discovery provides us with the exciting opportunity to better constrain the mass-luminosity relation for low-mass stars in the super metal-rich domain, expanding our understanding of the most-common types of stars and substellar objects.

\end{abstract}


\section{Introduction}

Binary stars are of great importance in stellar astrophysics. As the two stars in a system are gravitationally bound, we can derive important parameters such as mass, luminosity, period, and radius from these, as well as metallicity from studying the stars' spectra, which can give information about the underlying stellar physics. Solar-type stars, themselves, are in a multiple star configuration at a rate of about 45 \% \citep{raghavan10}  for metallicities between -1.0 and +0.6 dex \citep{jenkins15}.
As solar dwarf stars are common and very well characterized, they make for ideal laboratories to study less understood stars, such as low-mass M dwarfs.

{There are methods to investigate binary systems with a low-mass component, such as direct imaging (DI).
}
At the current time, young stars ($<$ 1 Gyr) are being targeted by DI searches as they are shown to make for feasible targets to detect planets, since planets this young are comparatively hot and bright (\citealp{marley07}).
Successful DI detections of young planets have allowed us to study some nearby systems, like HR 8799 (\citealp{marois08}), $\beta$ Pictoris (\citealp{lagrange10}), 51 Eridani
(\citealp{macintosh15})
. As of yet, no planets have been discovered orbiting solar-age stars with DI, owing to the large contrast between the planet and star. Nevertheless, a number of old low-mass stars and brown dwarfs at that age have been imaged (e.g. \citealp{burningham09}; \citealp{crepp12}, \citealp{mace13}; \citealp{crepp14}; \citealp{crepp16};
\citealp{ryu16}). %

Radial velocity (RV) surveys have been popular in recent years due to their productivity in discovering extrasolar planets in the Earth to Jupiter-mass regime at close separation to the star. One disadvantage to the RV method is that it is difficult to constrain orbits where the orbital period of the Doppler source is greater than the time baseline of the observations. We call these RV detections long-period trends. Linear trends are RV trends where we have little or no indication of any inflection (therefore appear linear) and can extract little of the source's orbital parameters. Another characteristic of the RV method is that the constrained orbits provide a minimum mass as $M \sin{i}$, not the absolute companion mass unless another method is used in conjunction with the RVs, like transit observations.

By combining RVs and DI, we can use the advantages of both methods to gain much more information of a star's orbital companion than that which can be measured by each method individually. Long-period RV trends make for good targets for DI searches as their probable large separation between the companion and primary bias the sample to where DI is most productive, at a distant separation from the star's bright halo. By tracking the orbits over time, the inclination can be derived in the images,
providing a true dynamical mass for the companion. 

{ For M dwarfs orbiting G dwarfs, we can use the masses from G stars' evolutionary models, along with the RV and DI parameters, in the derivation of the M dwarfs' dynamical masses. As well, we can use derived metallicities and ages from the more massive stars' models along with the dynamical mass and use them to constrain M dwarf evolutionary models, as the stars can be assumed to have been formed concurrently. }
{ We can make statistics for RV trends and discover if they relate to a companion of stellar-mass or substellar-mass, providing information about the formation mechanism of these low-mass companions. As we would obtain information about the frequency of M dwarfs orbiting G dwarfs, we gain a further understanding of how often M dwarfs are contaminants in planet searches, such as their being false positive eclipsing binaries in planet transit searches \citep{almenara09}. Imaging non-detections are also very useful, as an upper mass limit can be given for the orbital companion which would rule out a massive stellar companion \citep{guenther05}. These targets would be useful for future instrumentation to directly detect the brown dwarf or planetary-mass companion.}

{

Currently, the search for small planets around M dwarfs is becoming ever more popular, as highlighted by recent discoveries of planets orbiting { GJ1214 \citep{charbonneau09},} Proxima Centauri (\citealp{anglada-escude16}), TRAPPIST-1 (\citealp{gillon17}), { LHS1140 \citep{dittmann17}, and NGTS-1 \citep{bayliss17arxiv}}.
M dwarfs make for excellent targets to search for important orbiting planets as their habitable zones (HZs) are located at short distances from their surfaces, with only a relatively small difference in brightness compared to more massive Sun-like stars. The mass, radius, and density of detected planets can be inferred from the stars they are orbiting.  To disentangle these stars' characteristics and obtain precise masses for planets, it is important to measure well their luminosities and masses to characterize the population mass-luminosity relation, as has been done previously by \citet{henry93}, \citet{delfosse00}, and \citet{benedict16}, for example.

}

The Cal\'{a}n-Hertfordshire Extrasolar Planet Search (CHEPS; \citealp{jenkins09}) and EXoPlanets aRound Evolved StarS (EXPRESS; \citealp{jones11}) surveys have provided us with a wealth of candidates to follow-up with DI methods in the southern hemisphere. The CHEPS sample comprises a subset of metal-rich ([Fe/H] $\geq$ 0.1 dex) stars that have colors corresponding to late-F to mid-K spectral type (0.5 $\leq$ $B - V$ $\leq$ 0.9), are located in the Southern Hemisphere ($\delta \le$ 0$^o$), and exhibit low activity levels (log $R'_\mathrm{HK}$ $\leq$ -4.5). Metal-rich stars were chosen as they have been shown to have an increased probability of possessing giant planets (e.g. \citealp{fischer05}; \citealp{sousa11}).  Thus far the CHEPS has uncovered 14 new companions with masses ranging from that of Uranus up to the brown dwarf regime (\citealp{jenkins09};  \citealp{jenkins13b}; \citealp{jenkins17}). The EXPRESS sample was selected to be bright and nearby G and K (0.8 $\leq$ $B - V$ $\leq$ 1.2) giant stars (-0.5 $\leq$ $M_\mathrm{V}$ $\leq$ 4.0) in the Southern Hemisphere. These stars are shown to exhibit correlations between stellar mass, up to 2.5 $M_{\odot}$, and metallicity with the occurrence rate of giant planets \citep{jones16}.  The EXPRESS sample has currently discovered 11 substellar companions (\citealp{jones13}; \citealp{jones15a}; \citealp{jones15b}; \citealp{jones16}; \citealp{jones17}) and 24 spectroscopic binaries (\citealp{bluhm16}). Any DI detections from these two samples will provide important evidence for the formation mechanism of low-mass companions to metal-rich solar type dwarfs and giants.
{ The detection and constraint of orbits for gravitational companions will follow up the study by \citet{bonfils05} where the mass-luminosity relation for low-mass stars was considered along with the effect from metallicity.}

To date, there have been a small number of programs following up long-period radial velocity trends with DI, such as those of \citet{crepp12} and \citet{ryu16}, which show detections of "benchmark" low-mass companions that will be able to constrain dynamical masses in the future.
 
In this article, we introduce our SPHERE Ao Follow-up of Additional Radial velocity companIons (SAFARI) with our discovery of a 
low-mass stellar companion to HD 86006, a star with a linear trend from the CHEPS program. We discuss our selection of the two targets for this program, our observations, and our reduction procedure in Section~\ref{sec:reduction}. We show our estimation of the companion's fundamental characteristics, such as mass and temperature, as well as our constraints on the orbital parameters of the system in Section~\ref{sec:results}.
We also describe our observations of the star HD 90520 that resulted in a non-detection, and describe the lower-mass limit that was achieved in the same section. Lastly, we show our conclusions in Section~\ref{sec:conclusions}.

\begin{table}
\caption{Stellar characteristics of the stars HD 86006 and HD 90520.}
\center
\begin{tabular}{cccc}
\hline
\hline
Property & HD 86006 & HD 90520 & Reference \\
\hline
RA (J2000) &  09:54:31.169 &  10:26:10.648 & 1 \\
Dec. (J2000) & -45:43:51.53 & -45:33:45.16 & 1 \\
P.M. RA (mas/yr) & 53.388 $\pm$ 0.056 &  -85.150 $\pm$ 0.046 & 2 \\
P.M. Dec. (mas/yr) & 38.012 $\pm$ 0.054  &  79.833 $\pm$ 0.045 & 2 \\
V (mag) & 8.20 & 7.50 & 3 \\
J (mag) & 6.940 $\pm$ 0.021 & 6.412 $\pm$ 0.021 & 4 \\
H (mag) & 6.705 $\pm$ 0.051 & 6.148 $\pm$ 0.038 & 4 \\
K (mag) & 6.582 $\pm$ 0.017 & 6.064 $\pm$ 0.021 & 4 \\
Distance {(pc)} & $78.13 \substack{+1.55 \\ -1.49}$ & $64.02 \substack{+0.96 \\ -0.93}$ & 2 \\
Sp. Type & G5IV/V & G0IV/V & 5 \\

{[Fe/H]} & 0.33 $\pm$ 0.06 & 0.21 $\pm$ 0.07 & 6 \\
{[$\alpha$/Fe]} & 0.39 & 0.24 & 6\\
T$_\mathrm{eff}$ (K) & 5821 & 5946 & 6\\
Mass ($M_{\odot}$) & 1.27 $\pm$ 0.04 & 1.30 $\pm$ 0.05 & 6\\
log g (cm/s$^2$) & 4.129 $\pm$ 0.213 & 4.108 $\pm$ 0.244 & 6\\
Age (Gyr) & 3.67 $\pm$ 0.39 & 2.92 $\pm$ 0.46 & 6\\  
log($R'_\mathrm{HK}$) & -5.05 & -5.00 & 7\\
vsin(i) (km/s) & 3.4 & 4.6 & 7\\
Age (Gyr) & 7.95 $\pm$ 0.90 & 3.43 $\pm$ 0.89 & 8\\

\hline
\label{tab:values}
\end{tabular}

1: \citet{vanleeuwen07} \\
2: \citet{gaia16} \\
3: \citet{egret92}\\
4: \citet{cutri03} \\
5: \citet{houk78} \\
6: { \citet{soto18}}\\
7: \citet{jenkins11a} \\
8: \citet{casagrande11} \\
\end{table}

   \begin{figure}
   \center{\includegraphics[width=8cm,angle=0]{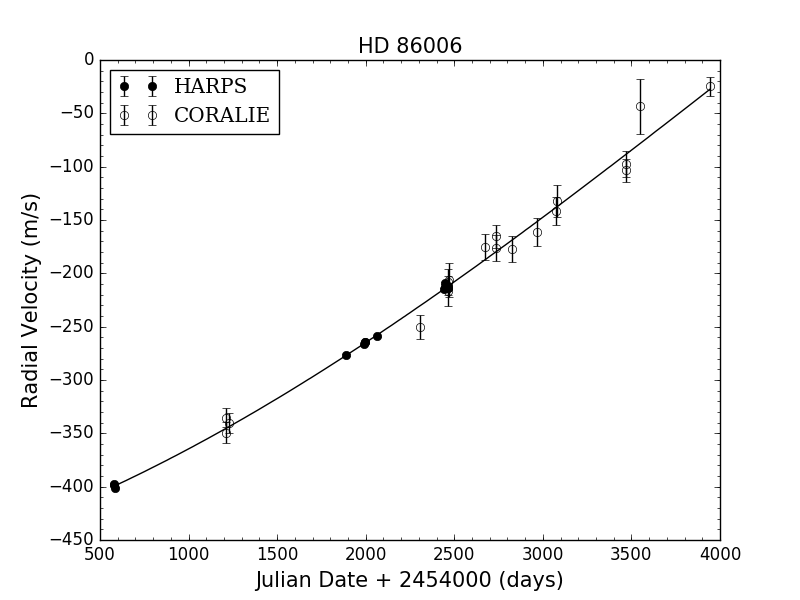}}
   \caption{Plot of radial velocity values for HD 86006. The closed data
   points represent data from HARPS, while the open points represent data from
   CORALIE. Note the linear trend without inflections making orbital
   characteristics difficult to determine.
}
      \label{fig:rvs1}
   \end{figure}

   \begin{figure}
   \center{\includegraphics[width=8cm,angle=0]{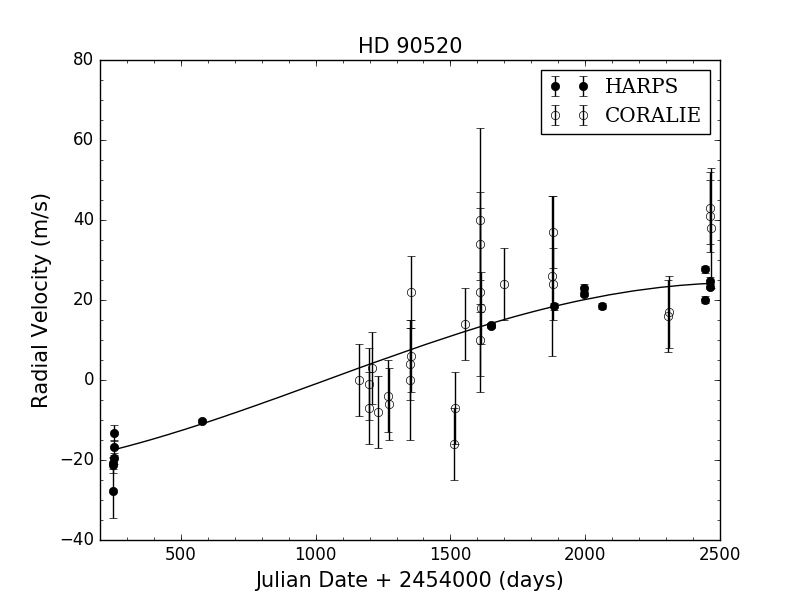}}
   \caption{Plot of radial velocity values for HD 86006. The closed data
   points represent data from HARPS, while the open points represent data from
   CORALIE. Note the linear trend without inflections making orbital
   characteristics difficult to determine.
   { %
}
}
      \label{fig:rvs2}
   \end{figure}

\begin{table}
\caption{Radial velocity data for HD 86006.}
\center
\begin{tabular}{cccc}
\hline
\hline
JD & RV (m/s) * & $\sigma$RV & Instrument \\
\hline
2454579.68825 & -125.73 & 0.91 & HARPS \\
2454580.65361 & -124.66 & 0.84 & HARPS \\
2454581.64833 & -127.95 & 0.66 & HARPS \\
2455883.87993 & -101.68 & 0.67 & HARPS \\
2455885.79942 & -3.74 & 0.86 & HARPS \\
\hline
\label{tab:rvs1}
\end{tabular}
\begin{tablenotes}
{ * Note that RV is a modified RV. This  is the RV measured by the instrument subtracted by the mean RV and secular acceleration.

{
** This table is an abbreviated version of the full table that is available as supplementary material in the online version of the publication.}}
\end{tablenotes}
\end{table}

\begin{table}
\caption{Radial velocity data for HD 90520.}
\center
\begin{tabular}{cccc}
\hline
\hline
JD & RV (m/s) * & $\sigma$RV & Instrument \\
\hline
2454248.45544 & -32.41 & 1.09 & HARPS \\
2454248.45899 & -33.63 & 1.05 & HARPS\\
2454248.46261 & -31.56 & 1.05 & HARPS\\
2454249.45329 & -27.75 & 6.82 & HARPS\\
2454250.45131 & -26.85 & 1.83 & HARPS\\

\hline
\label{tab:rvs2}
\end{tabular}
\begin{tablenotes}
{ * Note that RV is a modified RV. This  is the RV measured by the instrument subtracted by the mean RV and secular acceleration.

{
** This table is an abbreviated version of the full table that is available as supplementary material in the online version of the publication.}}
\end{tablenotes}
\end{table}

\section{Target Selection, Observations and Reduction}
\label{sec:reduction}

The targets we selected for SPHERE Science Verification (SV) time were taken from the CHEPS (\citealp{jenkins09}) and  EXPRESS projects \citep{jones11}. From the CHEPS program of about 150 metal-rich F to K stars stars and EXPRESS program of about 170 giant stars, the targets were chosen from a subset of 19 stars from CHEPS and 14 stars from EXPRESS that were shown to exhibit long period trends in their radial velocities, indicating they were hosts to very long period companions.  Although some of our selected sample exhibited significant curvature in their velocity timeseries, we decided that stars with linear trends were preferable in this case since the companions would be widely separated from the star, and given that this was a SV program where the instrument was being tested, the largest possible star-companion separation would maximise the possibility that we would obtain a positive result. HD 86006 and HD 90520, both from CHEPS, made for ideal targets due to their large linear RV trends and favourable sky positions for SV time, and in Table~\ref{tab:values}, we list their stellar characteristics. We are interested in targets that display some curvature as well and are including them for the next observations in our survey.

 The [Fe/H], [$\alpha$/H], $T_{\mathrm eff}$, mass, log g, and age parameters were obtained using the SPECIES code \citep{soto18}. SPECIES measures the FeI and FeII lines from high-resolution echelle spectra. By applying the ATLAS9 model atmospheres \citep{castelli04} and local thermodynamic equilibrium, it can obtain the surface gravity, metallicity, and temperature. The stellar photosphere's microturbulence velocity and the stellar rotational velocity are also obtained by using synthetic spectra to fit the absorption lines along with temperature relations. From the Isochrones \citep{morton15} package that interpolates the MIST Isochrones \citep{dotter2016}, the mass, radius, and age are obtained.
Finally for these stars, $\log{R'_\mathrm{HK}}$ and $v \sin{i}$ were obtained from \citet{jenkins11a}.

\subsection{Radial Velocity Observations}

The radial velocity data for these stars were observed by CORALIE at the Swiss Leonhart Euler Telescope (\citealp{queloz00}) and using the High Accuracy Radial Velocity Planet Searcher (HARPS) at the 3.6m Telescope (\citealp{mayor03b}) at the La Silla Observatory in Chile. The observations of HD 86006 were made over a timespan of nine years, running from April 2008 to { July 2017}. For HD 90520 they were made over six year period, from May 2007 to June 2013. 
 
CORALIE is an echelle fibre-fed spectrograph with a spectral resolving power of $\sim$50,000 covering the wavelength range between 3900-6800 \AA. It regularly maintains a radial velocity precision of better than 10 m/s. The CORALIE data was reduced using the normal steps for echelle spectra, including debiasing the images, locating and measuring orders with polynomial fitting, flatfielding, removing scattered light, making a 2D barycentric corrected wavelength solution, cross-correlating the spectra with a binary mask suitable for the spectral type, and fitting the cross-correlation function to calculate a radial velocity. Finally, the instrumental drift was measured with a simultaneous observation of a Thorium-Argon lamp, and this drift was then subtracted from the velocity measurement. For this process we used the CERES pipeline package (\citealp{brahm17}).

HARPS is an even more powerful radial velocity machine, since this echelle spectrograph has a resolving power of $\sim$115,000 and is heavily optimized for thermal, pressure, and mechanical stability by being housed in a mechanically stable vacuum chamber to negate the effects of air pressure and temperature variations on the instrument. The HARPS data was reduced using the HARPS-DRS following the procedure outlined in \citet{baranne96} and \citet{pepe02}, which follows similar steps to those outlines above for CORALIE. Long-term stability for the instrument has been demonstrated to be $\sim$ 1 m/s or better (see \citealp{locurto10}). More details of the radial velocity observations, pipeline processing, and computations for the CHEPS datasets can be found in \citet{jenkins17}. 
{ For our observations for both stars, we used the G2 binary mask to cross-correlate with the observed echelle spectra. For both CORALIE and HARPS observations, the majority of the observations were taken with a S/N of 50-100 at 5500 \AA.}

From CHEPS observations we have obtained 35 radial velocity observations using HARPS and CORALIE over 9.2 years for HD 86006 as shown in Table~\ref{tab:rvs1}. Using the Systemic Console (\citealp{meschiari09}), we fit a long-period trend to the data, as shown in Fig.~\ref{fig:rvs1}, where the filled circles are HARPS data and the open circles are CORALIE observations.  We repeated the same exercise for the star HD 90520, shown in Fig.~\ref{fig:rvs2}, where we have obtained 61 observations with the same instruments for this star, with a full baseline of over 6 years. 
While there may be some evidence for curvature in this data, a linear fit is also a viable solution, even though only an extreme lower limit can be placed on the companion period and minimum mass. We do not report the physical parameters of these solutions as they were only made to provide us with the orbital separation lower limit with a fixed eccentricity of 0, 
which are 16.8~AU and 6.7~AU for HD 86006 and HD 90520,
respectively, and the fits return reduced $\chi^2$ statistics of 3.6 and 3.4,
respectively after removing outliers in the residuals. Therefore, HD 86006 and HD 90520 became ideal targets for observation in SV with SPHERE due to these long-period trends, their physical brightnesses, and their sky coordinates. Finally, we show the Julian dates, radial velocity measurements, measurement uncertainties, and the instrument used in Table~\ref{tab:rvs1} for HD 86006 and Table~\ref{tab:rvs2} for HD 90520. {The discrepancy between the values shown in the Figs. \ref{fig:rvs1} and \ref{fig:rvs2} and Tables \ref{tab:rvs1} and \ref{tab:rvs2} are explained by instrumental offsets determined for the CORALIE and HARPS data sets, used to combine the data into one set and best fit an RV trend.}

\subsection{Direct Imaging Observations}

\subsubsection{SPHERE Observations}

We obtained observational imaging data of our targets HD 86006 and HD 90520 on
January 4, 2015, and February 22 and 23, 2015 (SV time), during ESO Period 97 on April 6, 2016, and during ESO Period 98 on November 14, 2017, with long-slit spectroscopic data taken on April 6, 2016 and May 2, 2016 on SPHERE \citep{beuzit08} %
 on the VLT UT3 at Cerro Paranal in Chile (ESO Run ID: 60.A-9385(A), 097.C-0775(A)).
We used SPHERE in IRDIFS mode, a mode making simultaneous observations with
the Infrared Dual-band Imager and Spectrograph (IRDIS, \citealp{dohlen08}) and the Integral
Field Spectrograph (IFS, \citealp{claudi08}). We used the H23 dual-band filters (1583 nm, 1667 nm) for IRDIS (\citealp{vigan10}) 
and IFS in the Y-J part of the spectrum ($R \sim$ 50, 950 nm - 1350 nm). We used the long-slit spectrograph (LSS) mode to follow up with a medium-resolution ($R \sim$ 350) of HD 86006 (\citealp{vigan08}). Our SV observations were made with the four-quadrant
phase mask (4QPM), while the P97 and P98 ones were taken with the apodized Lyot coronagraph (ALC) as the 4QPM is no longer offered. As the January 4, 2015 data was not centered properly and the LSS data of April 6, 2016 did not have the companion in the image, we have not used them in the final analysis.

All SPHERE IRDIS observations were taken with flux calibration images that were done by %
offsetting the PSF core so that the coronagraph is not covering it, which allows one to accurately measure the star's photometry, and also with star
center calibrations that allow accurate measurements of the star's position for all images, which is done by making four symmetric satellite spots on the coronagraphic image where the intersection is the location of the star under the coronagraph. The flux observations used neutral density filters to lessen the light and not immediately saturate the detector. 

The SPHERE IRDIS data was reduced in the typical way using the SPHERE Pipeline Recipes (v. 0.15.0, \citealp{pavlov08}) doing a background subtraction, flatfield division, star center determination, and deletion of bad pixels. The IRDIS detector does have a small anamorphic distortion, but as the distortion map calibration in the SPHERE pipeline does not report the appropriate mask offset, we use the distortion correction of \citet{maire16} by multiplying the Y coordinate by 1.0062 $\pm$ 0.0002. 
The SV data for HD 90520 was processed with angular differential imaging (\citealp{marois06}) and principal component analysis (\citealp{soummer12}) to be able to provide the best subtraction of the surrounding speckle halo and obtain the best possible contrast between a companion and its host star.

We used the centering calibration to measure the position of the primary star, as described above, {for the SPHERE observations on Feb. 12, 2015}. To obtain the position of the secondary star, we measured the centroid of a 2D Gaussian fit to the PSF with a full-width at half maximum (FWHM) described by that of the primary star in the flux calibration frame. To obtain a measurement of the separation between the star and the companion we used the pixel scale measured by \citet{maire16} as well as the parallax for the distance to provide us an absolute separation in AU, and we used their true north that was obtained using observations at a measurement date very close to our own to obtain the proper position angle.

To obtain an uncertainty on the centroid measurement, we rotated and derotated each frame by 5 degrees multiplied by a random number from a normal distribution 10 separate times and used the standard deviation on the mean measurement. {We then propagated the errors for each centroid measurement for each frame observed, as the frames were stacked.}
We propagated this error with the error on the detector distortion and the centroid measurements of the four satellite spots. We also propagated in the pixel scale error from \citet{maire16} to obtain an error on the separation between the primary and the companion. To obtain an error on the position angle, we propagated the angle measured by the centroid as done for the separation and error on the true north as measured in \citet{maire16}.

{For the same Feb. 12, 2015 data,}
we measured the flux of the primary star in the flux-calibration image by summing the detector counts for pixels within 1 FWHM of the centroid, correcting for the exposure time for the frame. We also corrected for the neutral density filter in place (SPHERE filter ND2.0 in this case, \citealp{vigan10}). To measure the flux of the companion, we measured the counts within 1 FWHM of the companion centroid in a science frame. We corrected for the exposure time and the speckle contribution by subtracting the flux using 1 FWHM apertures at the same distance as the companion surrounding the primary star under the coronagraph and took the mean of these values. These measurements provided us with the photometric contrast between the primary and the companion. 

For SPHERE observations other than the one taken on Feb. 12, 2015, we measured the astrometry and photometry using the flux-calibration image, as the companion is visible in that frame but the coronagraphic images had the companion saturated. 
The separation and position angle was measured by finding the distance and angle between the centroids, respectively, taking the parallactic angle of the observation into account. For SPHERE, the plate scale for non-coronagraphic images and true north for the proper period of time were used from \citet{maire16}. 
{ We used the Vortex Image Processing (VIP) package  (\citealp{gomezgonzalez16}). to implement our post-processing of our images, as described above.}

\subsubsection{MagAO Observations}

We also made an observation of HD 86006 using the Magellan Adaptive Optics (MagAO) instrument (\citealp{close08}) at the 6.5 m Clay Telescope at Las Campanas Observatory in Chile with the Clio2 infrared detector on November 26, 2015. We made observations in the H and K bands with the narrow camera (pixel scale of $\sim$ 15.85 mas pixels). It should be noted that the H filter is read noise limited, rather than sky limited, with 70 electrons of read noise. The K band is sky limited. The weather conditions were not ideal during that night.

Our MagAO data was taken with an ABBA nodding pattern. We reduced the data by first subtracting each A frame by a B frame taken closest in time in the same mode to subtract the background and dark current. We then stacked the detector integrations and aligned the nods. 
We stacked the images after centering each image from a fit of the PSF with a Gaussian function.
We did not apply any flatfield division, since there is out-of-focus image that is unremoveable by flat-correction (\citealp{morzinski15}).

To calibrate MagAO's astrometry, we used a MagAO observation of 47 Tucanae that we compared with one from HST, effectively allowing us to obtain a pixel scale and true north for this instrument; we describe this in more detail in Section \ref{sec:hst}.
{
We propagate the astrometric error by considering the standard deviation of the centroids from separate clean observed images, along with the pixel scale error, and the standard deviation of the position angles between the two centroids, along with the true north error. }

To measure the photometry, we measured the flux within 1 FWHM aperture of the primary star and the flux within 1 FWHM of the companion star. As with the SPHERE coronagraphic image, we subtracted the speckle background within the mean of apertures of the same radius and distance as the companion surrounding the primary in the center. The photometry error was measured by considering that photon counts follow a Poisson distribution, and therefore we estimate the error as $\sqrt{e^- * gain}$.

{
We also used the VIP package \citep{gomezgonzalez16} to implement the MagAO image post-processing.}
We summarize the observation dates, exposure times and modes in Table~\ref{tab:obs}.

\begin{table*}

\caption{Observations Table}
\center
\resizebox{\textwidth}{!}{
\begin{tabular}{cccccccccc}
\hline
\hline
Instrument & Mode & Date (DD/MM/YYYY) & Universal Time & JD & Object & Filter & No. of Exposures & Exposure Time (sec) & Calibrations \\
\hline
HST-WFC3 & IR & 18/03/2014 & 10:48:03 & 2456734.95004214 & 47 Tuc & F110W & 1 & 99 & \\
SPHERE & IRDIFS & 12/02/2015 & 03:51:07.591 & 2457065.66050452 & HD 86006 & H23 & 16x20 & 8 & Star Center, Flux Calib. \\
SPHERE & IRDIFS & 13/02/2015 & 03:16:42.295 & 2457066.63660064 & HD 90520 & H23 &  8x20 & 16 & Star Center, Flux Calib. \\
MagAO & Clio2 & 26/11/2015 & 02:15:45 & 2457352.59427  & 47 Tuc (center) & Ks & 20x2 & 1500 & AB pattern \\
MagAO & Clio2 & 26/11/2015  & 07:33:05 &  2457352.81464 & HD 86006 & H & 20x4 & 1500 & ABBA pattern, 500 sec. Flux Imgs. \\
MagAO & Clio2 & 26/11/2015  & 07:41:14 & 2457352.82030 & HD 86006 & Ks & 10x8 & 3000 & ABBA pattern, 500 sec. Flux Imgs. \\
SPHERE & IRDIFS & 06/04/2016 & 00:49:46.578 &  2457484.5356687 & HD 86006 & H23 & 1 & 64 & Star Center, Flux Calib. \\
SPHERE & IRDIS-LSS & 02/05/2016 & 01:02:29.29.699 & 2457510.53287012 & HD 86006 & MR-YJH & 14 & 64 & Flux Calib. \\
SPHERE & IRDIFS & 14/11/2016 & 08:37:10.900 &  57706.3591539 & HD 86006 & H23 & 4 & 64 & Star Center, Flux Calib. \\
\hline

\label{tab:obs}
\end{tabular}
 }
\end{table*}

\subsection{Spectroscopic Reduction}

In Table~\ref{tab:obs}, we summarize the spectroscopic observations taken with SPHERE. To reduce the SPHERE IFS data, we used the pipeline from \citet{vigan15} and steps described in \citet{mesa15}.
This pipeline works by first subtracting the background frames and four sets of detector flats for the YJ mode which include those taken in white light, in 1020 nm, 1230 nm, and 1300 nm filters. Then the spectra positions were defined by a calibration with light illuminating the detector evenly through the instrument. To calibrate the positions of the wavelengths on the detector, three monochromatic lasers illuminated the detector through the system. The integral field unit (IFU) flat is used to correct the lenslet contribution to science images. The calibrations were processed using the SPHERE DRH Pipeline.

To analyze the SPHERE IFS data, we find the position coordinates of the companion source for each spectral channel by using the daofind module, an implementation from the DAOPHOT algorithm described in \citet{stetson87}, in Photutils v. 0.2.2 \citep{bradley16}. Using these coordinates, we measured the FWHMs in the x and y directions, and used the mean as the FWHM for the PSF in each channel. The photons from the source were counted within 2 FWHMs of the centroid of the secondary companion and any residual background left after the background subtraction was subtracted using counts within apertures of 8.5 and 11.5 FWHMs from the centroid.  

We reduced the SPHERE LSS coronagraphic data by using a reduction pipeline of \citet{vigan16}. The pipeline uses a combination of the standard SPHERE DRH recipes with custom IDL routines to provide reduced and aligned LSS spectra. The pipeline then offers different algorithms for the subtraction of the speckle pattern from the data before the spectrum of the companion can be extracted. For this object, we used an improved version of the method based on spectral differential imaging described in \citet{vigan08}  as well as a simple subtraction of the symmetric speckle halo with respect to the star. After the speckle subtraction, the spectrum of the companion is extracted using an aperture of size $\lambda/D$ in each of the spectral channels. The noise is estimated from an identical aperture located on a symmetric position with respect to the star. The spectrum of the companion is then normalised to the flux of the primary in each channel, extracted using a similar aperture. Since the companion is very bright compared to the stellar halo and speckles, the spectra extracted with the two speckle subtraction methods yielded completely equivalent results. Because IRDIS provides two identical fields of view in LSS mode, the spectra obtained in each field were combined with a mean to increase the signal-to-noise ratio of the companion spectrum.

{ To account for the telluric bands in the spectra, for both the LSS and the IFS, we divided the spectra extracted for the secondary companion by the primary PSF spectrum.}
Finally, the spectrum, calibrated in contrast, was multiplied by a Planck function at the primary star's temperature (5620 K for HD 86006) to compensate for its spectral slope { that was divided out in the telluric calibration.} %
We compare the resulting spectrum to models and empirical objects { in Section~\ref{sec:modelcomp}}.

\subsection{HST Astrometric Calibration}
\label{sec:hst}

{
47 Tuc at RA: 00:23:58.12, Dec: -72:05:30.19 (J2000) is one of the astronomical objects used by SPHERE to calibrate it for astrometric measurements (\citealp{maire16}). To compare the MagAO and SPHERE images, we made observations of the cluster to calibrate MagAO's astrometric measurements to that of SPHERE. MagAO's observation was taken at a different point in the cluster (RA: 00:24:05.36, Dec: -72:04:53.20, J2000) making it necessary to compare it to other calibrated observations such as those from the Hubble Space Telescope Wide Field Camera 3 (WFC3).  We used the AstroDrizzle-reduced .drz image to correct for the distortion in the WFC3 detector. 

We used two stars in the image and measured their separation in sky coordinates given from the frame. From this we could calibrate the pixel lengths to distances in arcsecs from the HST observation. We also calculated the position angle of one star relative to the other to obtain a true north for the MagAO system. This allows us to compare the observations with SPHERE as the measurements are then on an absolute scale. 

From this we obtained a pixel scale for MagAO of 15.764 $\pm$ 0.125 mas/pix and a true north of -2.403 $\pm$ 0.033 $\degree$.  
}

\section{Results}
\label{sec:results}

\begin{table*}
\label{tab:hd86006}
\caption{Photometry and Astrometry of HD 86006B}
\center
\resizebox{\textwidth}{!}{
\begin{tabular}{cccccccc}
\hline
\hline
Instrument & Mode & JD & Filter & Separation (mas) & Position Angle (deg) & $\Delta$mag\\
\hline

SPHERE & IRDIS & 2457065.66050  & H2 & 331.46 ${\pm}$ 2.51 & 301.20 ${\pm}$ 0.13  & 4.21  ${\pm}$ 0.04  &\\
SPHERE & IRDIS & 2457065.66050 & H3 & 331.53 ${\pm}$ 2.51  & 301.35 ${\pm}$ 0.13 & 4.12 $\pm$ 0.04  &\\
MagAO & Clio2 &  2457352.81464 & H & 319.69 $\pm$ 9.53 & 302.28 $\pm$ 2.23 & 3.78 $\pm$ 0.14 & \\
MagAO & Clio2 & 2457352.82030  & Ks & 314.49 ${\pm}$ 6.74 & 300.83 $\pm$ 1.26 &  3.82 $\pm$ 0.22 & \\
SPHERE & IRDIS  &  2457484.53567  & H2  & 319.79 $\pm$ 2.35 & 301.57 $\pm$ 0.20 & 4.20  ${\pm}$ 0.05  &\\
SPHERE & IRDIS  &  2457484.53567  & H3  & 320.33 $\pm$ 2.36 & 301.47 $\pm$ 0.20 & 4.13 ${\pm}$ 0.05  &\\
SPHERE & IRDIS  &  2457706.35915  & H2  & 313.33 $\pm$ 2.31 & 301.24 $\pm$ 0.19 & 4.01 $\pm$ 0.05 & \\
SPHERE & IRDIS  & 2457706.35915  & H3  & 313.55 $\pm$ 2.31 & 301.32 $\pm$ 0.20 & 4.00 $\pm$ 0.04 & \\
\hline
\end{tabular}
}
\end{table*}

\begin{table}
\caption{HD 86006 System Characteristics}
\center
\begin{tabular}{ccc}
\hline
\hline
Characteristic & HD86006A & HD86006B \\
\hline
Spectral Type &  G5IV/V  & {M3.7 ${\pm }$ 1.1} $^a$ \\
& & {M4.5 ${\pm }$ 1.8 $^b$} \\
& & M3V $^c$\\
\hline
$T_{{\mathrm eff}} (K) $ &  5700  & 3300 ${\pm }$ 100 $^d$\\
 &  & 3600 ${\pm }$ 100 $^e$\\
 &  & $3258 \substack{+201 \\ -180}$ $^f$ \\
 &  & $3127 \substack{+320 \\ -370}$ $^g$ \\

\hline

$^a$ H${_2}$OA index\\
$^b$ H${_2}$OC index\\
$^c$ Best fit SpeX spectrum to LSS \\
$^d$ Best fit BT-Settl model to LSS \\
$^e$ Best fit BT-Settl model to IFS \\
$^f$ H${_2}$OA index converted to $T_{{\mathrm eff}}$\\
$^g$ H${_2}$OC index converted to $T_{{\mathrm eff}}$\\
\label{tab:system}
\end{tabular}
\end{table}

\subsection{HD86006 Companion Detection}

\begin{figure*}
\center{\includegraphics[width=14cm,angle=0]{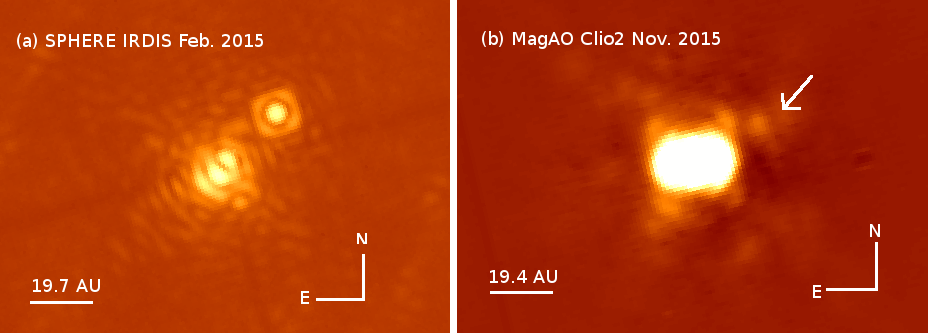}}
\caption{The left plot shows an image of HD 86006 with SPHERE IRDIS and the right is from MagAO. The M dwarf companion can be seen to the upper right of the star at a separation of $\sim$26 AU from the primary. }
\label{fig:hd86006_sphere}
\end{figure*}

\begin{figure}
\center{\includegraphics[width=9cm,angle=0]{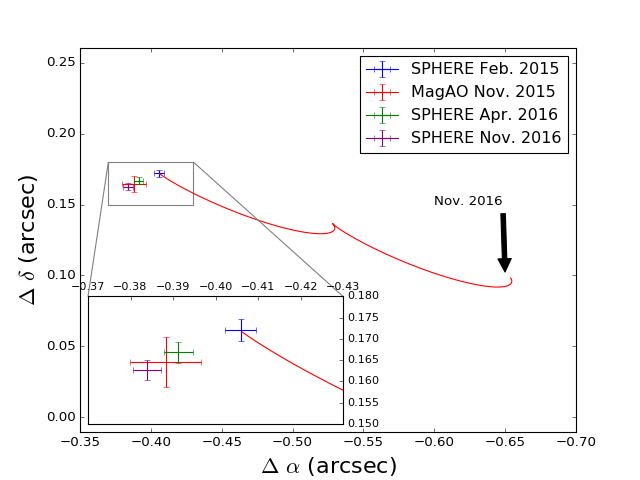}}
\caption{{Plot of the astrometric} positions of the companion to HD 86006 from our data with SPHERE and MagAO, along with a model trend of where the companion would be if it were a background source ending in November 2016 (red curve). The astrometric points are given for each epoch in different colors for the instrument and data, as shown in the legend. %
}
\label{fig:hd86006_pm}
\end{figure}

We made the detection of a companion to HD 86006 during the SV period as shown in Fig.~\ref{fig:hd86006_sphere}. This companion has an average contrast of $\Delta$H2 = 4.14 $\pm$ 0.03 {mag}, $\Delta$H3 = 4.08 $\pm$ 0.03 {mag} and a separation of 0.331 arcsec on the SV observation. We show the astrometric and photometric parameters and uncertainties in Table \ref{tab:hd86006}. The follow-up observations of the star with MagAO and SPHERE show that there is common proper motion as there is little change in the position for each image. In Fig.~\ref{fig:hd86006_pm}, we show the astrometric positions of the detected companion along with the track of where the companion would be if it were a background source moving slowly across the sky. As the companion's astrometric positions do not follow the track, the companion is gravitationally bound to its host star. The companion shows movement between the three SPHERE points and MagAO point.

\subsection{Spectral Indices}

We measured the temperature and spectral type of HD 86006B using empirically derived spectral indices. We considered the H${_2}$OA, H${_2}$OC indices from \citet{mclean03}, which were shown to well-represent early M stars in \citet{cushing05}.
These indices correspond to water absorption features and have a significant effect in low-mass stars. As these features are shown to relate to the spectral index of these low-mass stars, we can use empirical relations to derive a spectral index and from that a temperature of the companion. 
To calculate them, we used the LSS spectrum and took the median values for flux ranges in the reduced spectrum. 

We use the following relations to obtain a spectral type from \citet{mclean03}:\\

{\centering{
SpT = - 26.18  (H${_2}$OA) + 28.09\\
SpT = - 39.37  (H${_2}$OC) + 38.94\\
}}

The H${_2}$OA index corresponds to that of an {M 3.7 ${\pm }$ 1.1} and the H${_2}$OC index shows that of an {M 4.5 ${\pm }$ 1.8}. To convert these spectral types into ${T_\mathrm{eff}}$, 
we interpolated between M dwarf spectral types and corresponding  $T_\mathrm{eff}$ from Table 5 from \citet{pecaut13}. This gave
$3258 \substack{+201 \\ -180}$ K and $3127 \substack{+320 \\ -370}$ K
for the H${_2}$OA and H${_2}$OB indices respectively, with the precision based on the index precision, as shown in Table~\ref{tab:system}. 

\subsection{Model Comparison}
\label{sec:modelcomp}

To obtain model-derived characteristic values for the companion, we used the BT-Settl models \citep{allard12} using the solar abundance from \citet{asplund09}. [On BT-Settl webpage, this is under model group AGSS2009.]
 
\begin{figure}
º\center{\includegraphics[width=9cm,angle=0]{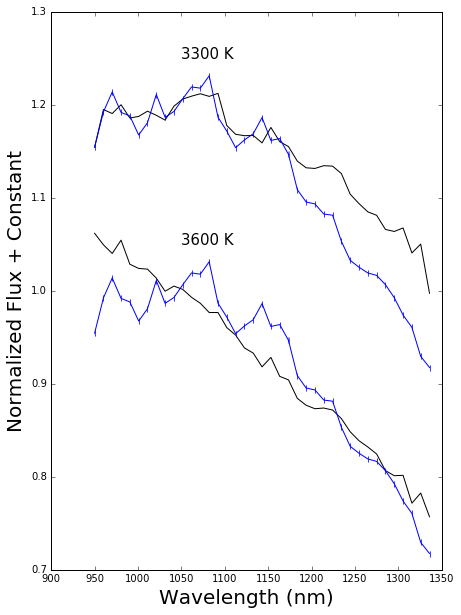}}
\caption{{The best fit 3600 K model (black) from the BT-Settl }to an SED from the IFS of HD 86006 (blue). This model also used a metallicity of +0.3. We also show a 3300 K, +0.5 metallicity model as it visually matches very well at the bluer wavelengths.}
\label{fig:hd86006_model}
\end{figure}

\begin{figure}
\center{\includegraphics[width=9cm,angle=0]{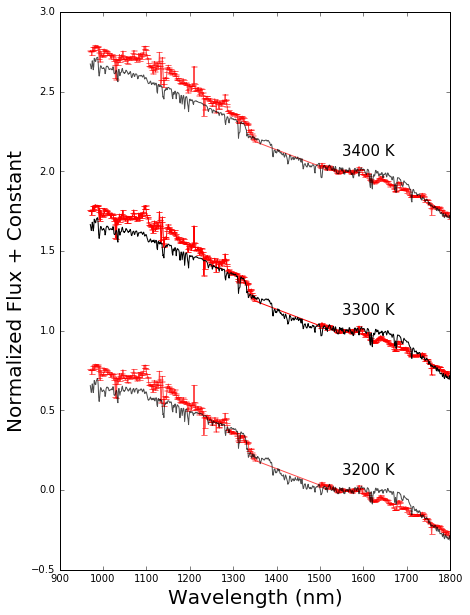}}
\caption{{The comparison BT-Settl spectra (3200 K, 3300 K, 3400 K) in black} with the LSS spectrum in red. The best fit using the maximum likelihood between all of the spectra was the one at 3300 K which also appears the best visually. For clarity, we have shown the spectrum with the water absorption feature removed.}
\label{fig:hd86006_lss}
\end{figure}

To see which BT-Settl model fits the observed spectra we obtained for the companion HD 86006B from the IFS and LSS of SPHERE, we convolved the model spectra to match the lower resolutions of the observations. We used a Gaussian maximum likelihood estimation of all models for low-mass stars with temperatures between 2600 and 5600 K, [Fe/H] either 0, 0.3, or 0.5 dex, and log(g) between -4.0 and -4.5 dex.
For the IFS spectrum (shown in Fig.~\ref{fig:hd86006_model}), the best fit obtained was 3600 K temperature given at a log(g) of -4.5 dex and [Fe/H] of +0.3 dex. In this, we see that the slope matches relatively well from 1150 nm to 1350 nm. At bluer wavelengths there is some divergence where the model rises higher than the observed SED. We also show in Fig.~\ref{fig:hd86006_model} the model for 3300 K, log(g) of -4.0 dex and [Fe/H] of +0.5 dex. Here visibly the SED for a high-metallicity cooler model matches very well in the blue from 950 nm to 1150 nm, but the slope diverges past 1150 nm to redder wavelengths. For the LSS spectrum, the maximum likelihood fit gave a 3300 K model with a log(g) of -4.5 dex and [Fe/H] of 0.0 dex as shown in Fig~\ref{fig:hd86006_lss}.{ We use 100 K uncertainties for our best fit, 
given that the spectra are provided in 100 K intervals.}

\begin{figure}
\center{\includegraphics[width=9cm,angle=0]{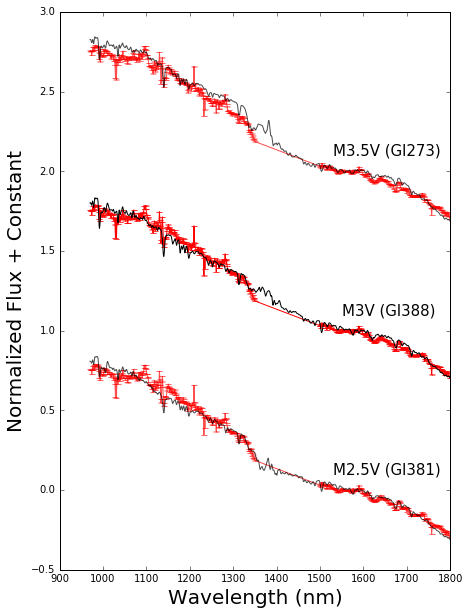}}
\caption{The LSS spectrum of HD86006B in red compared with the SpeX spectra in black. The Gl388 spectrum is visually a good fit to the LSS spectrum. An M2.5V (Gl381) and an M3.5V (Gl273) spectrum are also shown to compare.  For clarity, we have shown the spectrum with the water absorption feature removed.}
\label{fig:hd86006_spex}
\end{figure}

We also make use of stellar spectra from the NASA Infrared Telescope Facility (IRTF) Spectral Library (\citealp{cushing05}; \citealp{rayner09}) taken with the SpeX medium-resolution spectrograph at Mauna Kea. We made a visual comparison of the LSS spectrum with the spectra in the Library as shown in Fig.~\ref{fig:hd86006_spex}. The M3V spectrum of Gl 388 appears to provide the best fit
which has an effective temperature of 3390~K and [Fe/H] of 0.28 \citep{rojas-ayala12}. This agrees quite well with our measured effective temperature from H${_2}$O indices, along with the metallicity of the host star (+0.33 $\pm$ 0.06~dex; \citealp{soto18}).

To obtain physical parameters for the companion, we used the Isochrones code \citep{morton15}, written in Python, which makes use of MultiNest \citep{feroz09}, implemented as PyMultiNest \citep{buchner14}. 
To make estimates of the age and mass, we provided the code with isochrones from an evolutionary model and used the interpolation from the Isochrones code. We put many random points on the isochrone, cut the area that agreed with our magnitude and temperature observations, and used these points to estimate the mass and age.

\begin{figure}
\center{\includegraphics[width=9cm,angle=0]{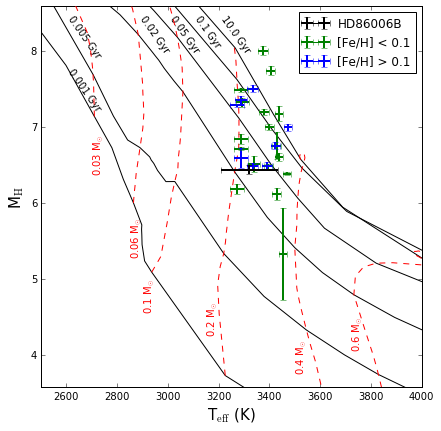}}
\caption{
The isochrones for BT-Settl models at masses varying from 0.03 M${_\odot}$ to 0.6  M${_\odot}$ and ages from 1 Myr to 12 Gyr as labeled. In black, we give the SPHERE H2 photometry of HD 86006B and the range of temperatures by taking the average of the temperatures found from SED fitting and spectral indices. We also show various solar neighborhood M3 dwarfs with their photometries and their temperatures and metallicities (for which we divide into two groups, with green representing low metallicity and blue representing high metallicity) from \citet{rojas-ayala12}. 
}
\label{fig:cont1}
\end{figure}

Using the AMES-Cond \citep{allard01, baraffe03} with [Fe/H] = 0.0 dex and BT-Settl \citep{allard12} with [Fe/H] = 0.3 dex model isochrones, we used the measured the SPHERE H2 and H3 photometry for the three epochs added with the \citet{cutri03} photometry for the primary star and derived temperature to find a mass and age for the companion. 
We note that the AMES-Cond model is recommended for objects with with  $T_\mathrm{eff}$ $<$ 1700, which is less than that of our discovered companion, while the BT-Settl model is considered valid from the stellar to planetary regimes. We use $T_\mathrm{eff}$ of 3321 $\pm$ 111 K, which was obtained by taking the mean of 
the $T_\mathrm{eff}$s for HD86006B found by the BT-Settl best fits for the IFS and LSS spectra (3600 K and 3300 K, respectively) and the two H$_{2}$O indeces \citet{mclean03}. From the BT-Settl isochrone, we obtain 0.23 $\substack{+0.10 \\ -0.04}$ M$_{\odot}$ for the mass and a range of ages between 15 and 59 Myr which we show in Fig.~\ref{fig:cont1}.
{For the AMES-Cond isochrone, we derive a mass of 0.23 $\substack{+0.09 \\ -0.07}$ M$_{\odot}$ and age between 12 and 53 Myr which is also below the ages derived from 
the primary star.
We show the obtained model isochrone parameters in Table~\ref{tab:models}.
}

{ We can use the age derived from the primary as a proxy age for the M dwarf as we can assume that they had formed together. This allows us to make a test on the isochrone-derived age of the companion, but it should be noted that ages tend to be poorly constrained parameters in general.}
{{The companion's} isochrone age estimate is very low, compared to the ages derived from the primary star 
(3.67 $\pm$ 0.39 Gyr and 7.95 $\pm$ 0.90 Gyr),} {which also has a low $\log{R'_\mathrm{HK}}$ of -5.05 dex suggesting an old age of Gyrs \citep{mamajek08}. This clear inconsistency between the measured ages for the M dwarf companion and the primary gives an indication that the models do not well constrain this parameter. }
To investigate further this discrepancy, we also plot the absolute H band magnitudes of solar neighborhood M3 stars with spectroscopically measured $T_\mathrm{eff}$s from \citet{rojas-ayala12} {in Fig. \ref{fig:cont1}}. Even though nearby stars, especially M dwarfs, tend to be old, on the order of Gyrs, their positions on the HR-diagram are scattered from $>$ 10 Gyr  down to $<$ 0.02 Gyr of age, with the majority found to be clustered around or below the 0.1 Gyr isochrone. This shows that currently the model isochrones for low-mass stars do not necessarily give precise age estimates, likely significantly underestimating the ages of mid-M stars. We also factor in the spectroscopic metallicities ([Fe/H]) from \citet{rojas-ayala12}, dividing the sample into two groups, with a dividing line set as 0.1~dex in metallicity to represent the metal-rich and metal-poor samples, and we see that there is no clear difference in the groups' positions on the isochrones, indicating that the shift towards a lower age estimate for HD 86006B is not due to its super metal-rich nature, as known from the primary star's [Fe/H] of 0.33 $\pm$ 0.06 {dex.}
{As the ages determined for HD 86006B by the isochrones is shown to be inconsistent, we are not assured that the measured masses are reliable.
}

\begin{table}
\caption{HD 86006B Model Physical Characteristics}
\center
\begin{tabular}{ccc}
\hline
\hline
Characteristic & AMES-Cond & BT-Settl \\

\hline

Mass ($M_{\odot}$) &  0.23 $\substack{+0.09 \\ -0.07}$ & 0.23 $\substack{+0.10 \\ -0.04}$ \\
log(Age {[yr]}) & 7.38 $\pm$ $\substack{+0.34 \\ -0.30}$ & 7.43 $\substack{+0.34 \\ -0.25}$\\

\hline
\end{tabular}
\label{tab:models}
\end{table}

\subsection{Model Fitting}

We developed a code in Python to make a joint analysis using a Markov-chain Monte Carlo method (MCMC) that combines the RV data and DI data, in order to best constrain an orbital solution. 
{ To do this, we developed a code that uses an orbital model based on Kepler's laws of planetary motion as given by the following Keplerian elements:}

$a$, semi-major axis,

$e$, eccentricity,

$i$, inclination,

$\Omega$, longitude of the node,

$\omega$, argument of periastron,

$\nu$, true anomaly at a given epoch

{ This Keplerian orbit is represented by a simple ellipse with the primary star being at a focus:}\\

$x = b \cos{\nu}$\\

$y = a \cos{\nu}$\\

$z = 0$\\

{ The distance from the center to a focus of the ellipse is given as}

$c = \sqrt{a^2 - b^2}$

{ The ellipse is multiplied by a rotation matrix to transform it into a new ellipse that represents a projection of the real ellipse on the sky.}

{
\[
\begin{bmatrix}
x'\\
y'\\
z'\\
\end{bmatrix}
=
R
\begin{bmatrix}
x\\
y\\
z\\
\end{bmatrix}
\]
}

where

{
\tiny{

\[
R =
\begin{bmatrix} 
\cos{\omega} \cos{\Omega} - \sin{\Omega} \sin{\omega} \cos{i} & \sin{\omega} \cos{\Omega} - \cos{\omega} \cos{i} \sin{\Omega} & \sin{\Omega} \sin{i}\\
\cos{\omega} \sin{\Omega} + \cos{\Omega} \sin{\omega} \cos{i} &  -\sin{\Omega} \sin{\omega} + \cos{\Omega} \cos{\omega} \cos{i} & -\cos{\Omega} \sin{i}\\
\sin{i} \sin{\omega} & \cos{\omega} \sin{i} & \cos{i}
\end{bmatrix}
\]
}}

$x'$ and $y'$ give the new coordinates for our projected orbit on the sky.

{ The semi-major axis, b, is defined as} \\

$b = a \sqrt{1 - e^2}$\\

As we want the orbit to be defined by the time (in Julian Date) of the astrometric points rather than by angle, we want to convert the angle (or true anomaly) to the time of periastron passage, $T_0$. We do this by converting the true anomaly, $\nu$, to the eccentric anomaly, $E$, and then the mean anomaly, $M$.\\

$\tan{E} = \dfrac{\sqrt{1 - e^2} \sin{\nu}}{1 + e \cos{\nu}}$ \\

$M = E - e \sin{E}$\\

We can use the mean anomaly, as it is the position angle of the companion in if it were on a circular orbit with the same period and speed, and the mean angular motion $n$ given by defined by a gravitational parameter, to find the time of its position, with $t_0$ being the time it passes periastron.\\

$n = \sqrt{\dfrac{G (M_1 + M_2)}{a^3}}$\\

$t = \dfrac{M + n t_0}{n}$\\

where $G$ is the universal gravitational constant, $M_1$ is the mass of the primary star, and $M_2$ is the mass of the secondary companion.

The radial velocity equation, { based on Kepler's third law of planetary motion,} that was used as part of our model to fit the radial velocity data, giving us the mass parameter and further constraining the orbital elements, is as follows:

{
\centering{
$v_r = v_0 + \sqrt{ \frac{G}{a_2 (M_1+M_2) (1-e^2)}}  M_2  \sin{i}  (\cos{\omega + \nu} + e \cos{\omega})$
}}

where ${v_r}$ is the radial velocity, ${v_0}$ is the barycentric velocity, $G$ is the universal gravitational constant, ${a_2}$ is the semi-major axis of the orbital companion, ${M_1}$ and ${M_2}$ are the primary and secondary masses respectively, $i$ is the inclination, $e$ is the eccentricity, $\omega$ is the argument of periapsis, and $\nu$ is the true anomaly. For the astrometric orbit, $\Omega$, the longitude of the node is also relevant.

Also we considered that the eccentricity and argument of periapsis are parameters that show a high degree of degeneracy, especially as the eccentricy approches 0, so we combine these two parameters to make new ones that are dependent on $e$ and $\omega$, as in \citet{marsh14}, with the following equations:\\

{\centering{
$x = \sqrt{e} \cos{\omega}$  and $y = \sqrt{e} \sin{\omega}$\\
$e = x^2 + y^2$\\
}}

We used a 2D Gaussian likelihood function with correlated parameters for the astrometric model and a 1D Gaussian likelihood function for the radial velocity model with consideration given to correlated noise.  To give a good starting point for the MCMC code to run, we used trial and error inputting different values for the parameters to get a close fit to the model and gave flat priors that were near these values. These "close" parameters with a small random value added were used as our starting values to run the code.  After running the chain with emcee \citep{foreman13}, we could obtain posterior probability distributions on the parameters. We used these distributions to obtain our estimations of the parameters at the 16th, 50th, and 84th percentiles which we used as their uncertainties.

\begin{figure*}
\center{\includegraphics[width=19cm,angle=0]{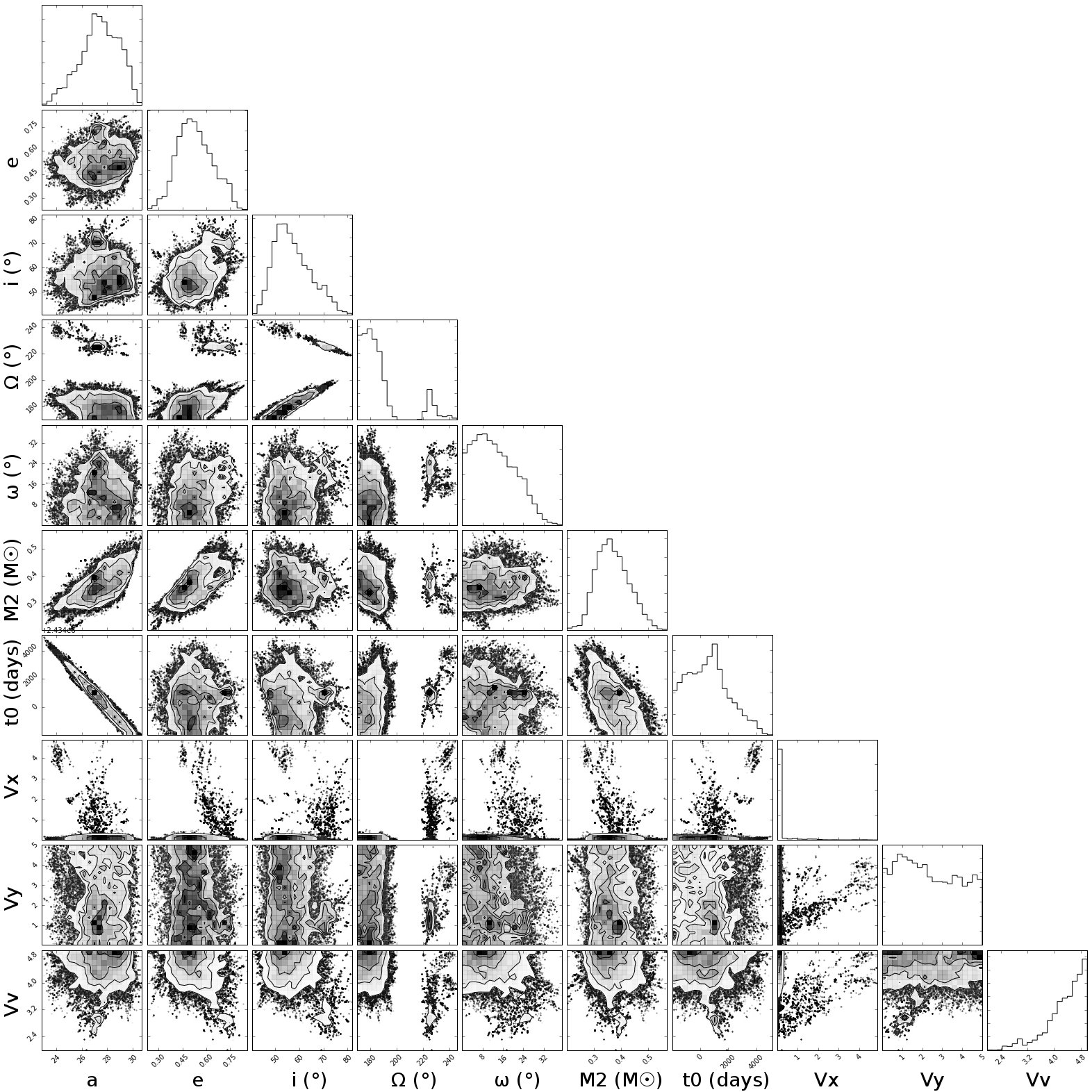}}

\caption{Corner plot for the MCMC astrometry and radial velocity fits. We show the 2D posterior probability distributions for each parameter with each of the others to demonstrate how well they correlate along with the 1D distribution for each at the top of their set of plots. Vx, Vy, and Vv are paramters representing the systematic noise or jitter for the x and y astrometric dimensions and the radial velocities, respectively. %
}
\label{fig:mcmc_corner}
\end{figure*}

\begin{figure}
\center{\includegraphics[width=9cm,angle=0]{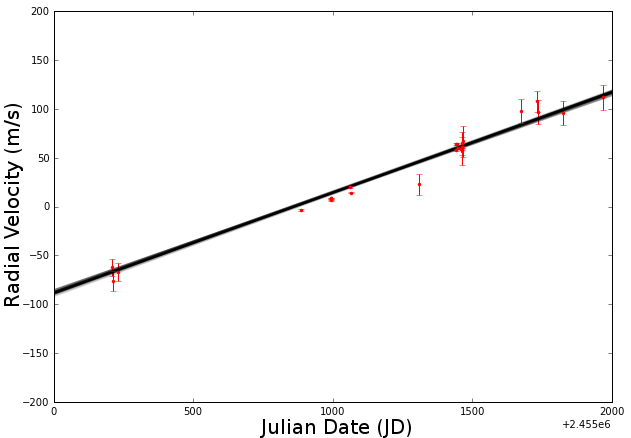}}
\caption{Plot of the fit from the MCMC code to the RV data. The red points shown with their error are the observed radial velocities and are the same as the points show in Table \ref{tab:rvs1}. The black curves represent one fit from the MCMC chain. As all fits after the burn-in were similar for the radial velocities, they nearly appear as a line in this plot.}
\label{fig:mcmc_rv}
\end{figure}

\begin{figure}
\center{\includegraphics[width=5cm,angle=0]{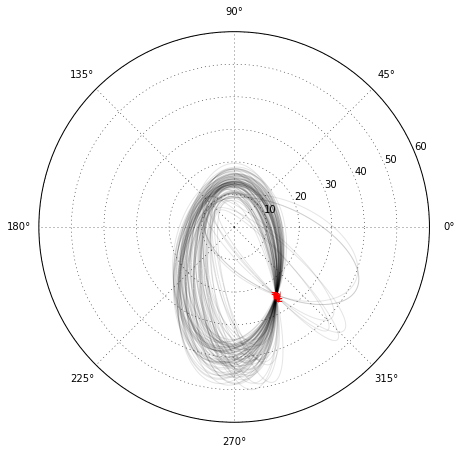}}
\caption{Fit to the astrometric points from SPHERE and MagAO from the MCMC code. The red points represent the position of observations, while each black line is the fit output from the MCMC run. The radial scale is in AU.}
\label{fig:mcmc_astrom}
\end{figure}

 When running the code using the uniform priors shown in Table~\ref{tab:keplerian}, we obtain a Keplerian fit for the companion to HD 86006 with a semi-major axis of $23.26 \substack{+1.75 \\ -1.80}$ AU, an eccentricity of $0.68 \substack{+0.12 \\ -0.12}$, mass of $0.321 \substack{+0.064 \\ -0.058}$ $M_{\odot}$, and other parameters shown in the same table. This mass is measured to 20$\%$, which is higher than the $<10\%$ mass measurements used in the mass-luminosity relation of \citet{delfosse00}, making more data necessary to further constrain the mass.
{Fig.~\ref{fig:mcmc_corner} shows the correlations each of the different parameters. We see correlation between the semi-major axis and  the time of periastron passage. We also see some possible correlation between a and $M_2$, i and $\Omega$, and $t_0$ and  $M_2$. Fig.~\ref{fig:mcmc_rv} shows the RV fit, and Fig.~\ref{fig:mcmc_astrom} shows the astrometric fit. We see that the fit is not very well constrained and will need to await more data before being able to make a confidently well-constrained dynamical mass.
}

\begin{table}
\caption{HD 86006B Keplerian Elements}
\center
\begin{tabular}{ccc}
\hline
\hline
Keplerian element & Prior & Value\\
\hline

a (AU) & $\mathcal{U}(10, 40)$ & $27.44 \substack{+1.54 \\ -1.57}$ \\
e & $^1$ &  $0.52 \substack{+0.13 \\ -0.10}$\\
i (\degree) & $\mathcal{U}(40, 89.74) + \mathcal{U}(90.26, 120) ^2$ & $56.13 \substack{+9.36 \\ -6.21}$ \\
$\Omega$ (\degree) & $\mathcal{U}(170, 250)$ &  $181.62 \substack{+10.85 \\ -7.80}$ \\
$\omega$ (\degree) & $^1$ & $11.80 \substack{+9.91 \\ -7.67}$\\
$t{_0}$(days) & $\mathcal{U}(2432000, 2442000)$ & $2434540 \substack{+1524 \\ -1590}$ \\
$M{_2}$ ($M_{\odot}$)  & $\mathcal{U}(0.2, 0.7)$ & $0.365 \substack{+0.067 \\ -0.055, }$\\

\hline

\end{tabular}
$^1$: e and $\omega$ are tied parameters as described in Section 3.5. The prior we set was a uniform distribution where $e^2 + \omega^2 < 1$\\
$^2$: The priors do not include a small area near 90 \degree as the code cannot calculate a completely edge on orbit.
\label{tab:keplerian}
\end{table}

\subsection{HD 90520}

\begin{figure}
\center{\includegraphics[width=9cm,angle=0]{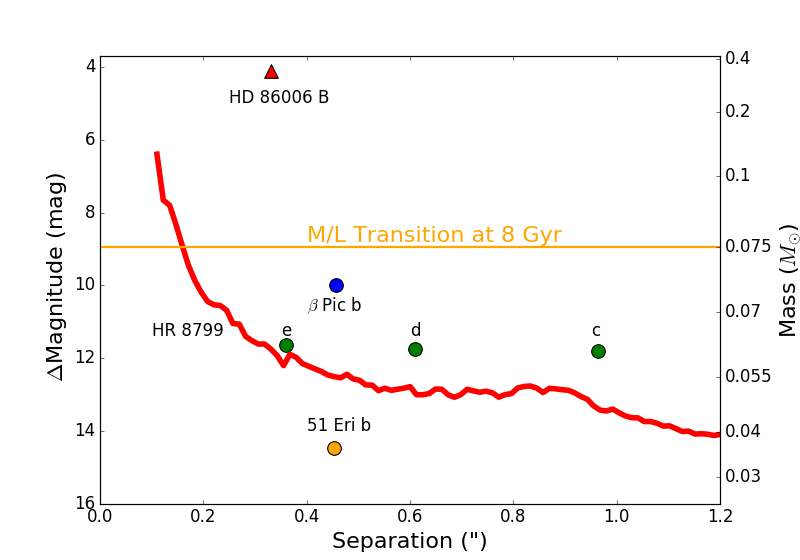}}
\caption{
{The 5$\sigma$ contrast performance curve over the separation from the SPHERE IRDIS SV observations of HD 90520. 
We also show the instrument's performance as a companion mass detectability limit (right axis) using the AMES-COND model to convert between magnitude and mass. We show where the directly imaged planets $\beta$ Pic b (\citealp{lagrange10}; \citealp{bonnefoy13}), HR 7899 bcde (\citealp{marois08}; \citealp{zurlo16}), and 51 Eri b \citep{macintosh15} would lie with respect to the contrast performance of these observations. We also highlight the position of our detection of HD 86006B and of the expected stellar to substellar transition for old objects.
}
}
\label{fig:hd90520contr}
\end{figure}

The SPHERE SV observations of HD 90520 did not produce a detection of a companion, but did provide us the best contrast performance for the SV run. We were able to place a strong constraint on the maximum mass and separation of the hidden companion using the contrast limit produced by an ADI-PCA method, along with accounting for small sample statistics close to the star (\citealp{mawet14})
The contrast curve is shown in Fig.~\ref{fig:hd90520contr}, which demonstrates SPHERE's sensitivity for this data set.  %
{Considering that the stellar SED agrees with that of spectral type of a
G0IV/V star,} 11 mag contrast in the H2 band at 200 mas corresponds to a maximum mass of 0.07 $M_{\odot}$ and a minimum mass detection of 0.06 $M_{\odot}$ at 500 mas.  This means that at these very close separations we can rule out any stellar companion giving rise to the trend we find in the RV measurements. { Therefore, this target must be a brown dwarf or planetary companion if it lies at a separation outside of 200 mas. From our RV simulation with the Systemic Console, we are able to fit a 2.2 $M_{Jup}$ planet with a 6.6 AU semi-major axis assuming an eccentricity of 0 and 90$\degree$ inclination at a reduced $\chi^2$ of 3.4 as shown in Fig.~\ref{fig:rvs2}. This will make for an ideal target to observe with future adaptive optics instrumentation.}

\section{Summary and Conclusions}
\label{sec:conclusions}

We used SPHERE SV in the H23 dual-filter IRDIFS mode to image directly the stars HD 86006 and HD 90520, chosen for their long-period linear radial velocity trends from the CHEPS survey. HD 86006 provided an M dwarf companion at about 25 AU from its primary which we have confirmed with following observations with MagAO and SPHERE. The IFS and LSS spectra were used to characterize the spectral type of the star with spectral indices and using the BT-Settl. 
The analysis of the LSS spectrum agreed with the expected spectral type from the contrast.

We used spectral fitting, spectral indices, and model isochrones to derive a mass, temperature, and age for the companion. The spectral fitting agreed with the star being M3.7 or M4.5, while our best fit template to the LSS spectrum agreed with that of an M3V star. The temperatures also agreed with that assessment, giving about 3300 K, or about M3-M4 in spectral type \citep{pecaut13}.
The ages derived from isochrone fitting were significantly lower than the expected age by over two orders of magnitude, and when comparing to nearby field M dwarfs with spectroscopically derived temperatures and metallicities, this was found to be the case also, indicating that theoretical models of mid-M dwarf evolution still require some fine tuning to match the observations. {Due to this, we do not assume the isochrone mass measurements to be reliable.}

We were able to also make a preliminary orbital solution for the data points taken with imaging. We will, over time, be able to map out the orbit of HD 86006B with imaging instruments, which will be able to provide a dynamical mass for the M dwarf. This discovery represents the first low-mass companion detection from our sample of long-period radial velocity trends from the CHEPS and EXPRESS samples.  As the stars in the CHEPS sample are intentionally biased to be high metallicity, and the EXPRESS sample also seems to favour metal-rich giants, our SAFARI project will prove to be fundamental to help constrain the mass-luminosity relation for metal-rich stars, of which there are few (\citealp{delfosse00}; \citealp{gaidos14}; \citealp{newton14}; \citealp{terrien15}). These new dynamical masses will then contribute heavily to gain a better understanding of the  physics of low-mass star formation, structure, and evolution.

\section{Acknowledgments}
BMP acknowledges support from CONICYT National Doctoral Scholarship Grant No. 21161783.
JSJ acknowledges support by Fondecyt grant 1161218 and partial support by CATA-Basal (PB06, CONICYT).
These observations are based on ESO Programs 60.A-9385, 097.C-0775,	098.C-0583, and 099.C-0878 as well as time allocated by the Chilean National Telescope Allocation Committee for use of the Magelland telescopes.
We would like to acknowledge the use of the SPHERE Data Center for
their help in the reduction of the SPHERE data. We would like to thank Pierre Kervella for providing his code for mapping the trajectory of a star based on proper motion and parallax.

\bibliographystyle{mnras}
\bibliography{refs}

\end{document}